\documentclass[11pt]{article}

\usepackage[]{hyperref}
\usepackage{fullpage}
\usepackage{amsmath,amssymb,amsthm,bbm,bm}
\usepackage{graphicx, booktabs, multirow, multicol, float}
\newcommand{\head}[1]{\textnormal{\textbf{#1}}}
\usepackage[toc,page, title,titletoc]{appendix}
\usepackage{authblk}

\usepackage{animacros}

\title{Bang-bang control as a design principle for classical and quantum optimization algorithms}
\author[1,2]{Aniruddha Bapat}
\author[3,4]{Stephen Jordan}
\affil[1]{Joint Center for Quantum Information and Computer Science, University of Maryland}
\affil[2]{Department of Physics, University of Maryland}
\affil[3]{Microsoft, Redmond, WA 98052, USA}
\affil[4]{University of Maryland, College Park, MD 20742, USA}
\date{}

\begin{document}
\maketitle

\begin{abstract}
  Physically motivated classical heuristic optimization algorithms such as simulated annealing (SA) treat the objective function as an energy landscape, and allow walkers to escape local minima. It has been argued that quantum properties such as tunneling may give quantum algorithms advantage in finding ground states of vast, rugged cost landscapes. Indeed, the Quantum Adiabatic Algorithm (QAO) and the recent Quantum Approximate Optimization Algorithm (QAOA) have shown promising results on various problem instances that are considered classically hard. Here, building on previous observations from \cite{mcclean2016, Yang2017}, we argue that the type of \emph{control} strategy used by the optimization algorithm may be crucial to its success. Along with SA, QAO, and QAOA, we define a new, bang-bang version of simulated annealing, BBSA, and study the performance of these algorithms on two well-studied problem instances from the literature. Both classically and quantumly, the successful control strategy is found to be bang-bang, exponentially outperforming the quasistatic analogues on the same instances. Lastly, we construct O(1)-depth QAOA protocols for a class of symmetric cost functions, and provide an accompanying physical picture.
\end{abstract}

\section{Introduction}

As quantum computing enters the so-called NISQ era \cite{nisq}, some focus has started shifting to noisy, shallow digital computations, and a need to re-examine existing quantum heuristic algorithms has emerged. The quantum adiabatic optimization algorithm (QAO), introduced in the previous decade \cite{Farhi2000}, provides a paradigm for quantum speedups in optimization problems, where one performs a quasistatic Schr\"{o}dinger evolution from an initial quantum state into the ground state of computational or physical interest. Runtime bounds for QAO typically depend, via adiabatic theorems, on the minimum spectral gap between the ground state and first excited state.

The Quantum Approximate Optimization Algorithm (QAOA) provides an alternative framework to designing quantum optimization algorithms, which is based on parameterized families of quantum circuits with adjustable parameters \cite{Farhi2014,hogg2000}. Such \emph{variational} circuits are parameterized by a depth, an initial quantum state, and a set of Hamiltonian operators under which the state can evolve. An instance of a variational circuit is further specified by a series of (labeled) evolution times that determine which operator is applied and for how long. Along with QAOA, several other recent models of heuristic computation fit into the variational circuit paradigm \cite{hadfield2017, Farhi2018, Kim2017, wecker2016, verdon2018}.

A primary distinguishing feature between the quasistatic paradigm of QAO and simulated annealing (SA) and the variational circuit paradigm is in the design of their evolution schedules, from quasistatic to a rapidly switching, or \emph{bang-bang}, schedule. Recently, it was observed \cite{mcclean2016, Yang2017} that the Pontryagin Minimum Principle \cite{pontryagin} implies that variational methods that employ a bang-bang evolution schedule are sufficient for optimality of the optimization protocol. Furthermore, the paper that introduces QAOA \cite{Farhi2014} also gives evidence pointing to an exponential speedup between QAOA and QAO. This raises two questions: Firstly, can a design shift from quasistatic to bang-bang yield provable superpolynomial improvements in the runtime, or are the two frameworks polynomially equivalent? Secondly, can the same control theoretic reasoning be applied to the design of classical optimization algorithms? In this work, we answer these questions by studying the performance of bang-bang controlled algorithms on certain well-studied instances, and make comparisons to the quasistatic, annealing-type algorithms. We prove that, on these instances, going from quasistatic scheduling to bang-bang can bring about an exponential speedup for both classical and quantum optimization. We also discuss the applicability and potential limitations of the optimal control framework to the problem of designing heuristic optimization algorithms.

\section{Summary of results}

The main results of this paper may be found in Sec.~\ref{sec:performance}, where we study the performance of four candidate algorithms given in Table~\ref{tab:algs} on two benchmarking instances, and find that the bang-bang control algorithms exponentially outperform both classical and quantum annealing-based algorithms. These results are also summarized in Table~\ref{tab:results}.

\begin{table}[H]
  \centering
  \begin{tabular}[h]{ccccccc}
    \toprule[1.5pt]
    \head{Instance} & \phantom{abc}&\multicolumn{2}{c}{\head{Annealing-based}} & \phantom{abc} & \multicolumn{2}{c}{\head{Bang-bang}}\\
     \cmidrule{3-4}\cmidrule{6-7}\head{}& & QAO & SA & &QAOA & BBSA\\
    \midrule[1.5pt]
    \texttt{Bush, $\lambda\ge 1$} & & poly$(n)$ \cite{Farhi2002} & exp$(n)$\cite{Farhi2002} & & $O(1)$\S~\ref{sec:QAOAbush} & $\tilde O\paren{n^{3.5...}}$\S~\ref{sec:BBSAbush} \\ 
    \texttt{Bush, $\lambda <1$} & & exp$(n)$ \cite{Farhi2002} & exp$(n)$\cite{Farhi2002} & & $O(1)$\S~\ref{sec:QAOAbush} & $\tilde O\paren{n^{3.5...}}$\S~\ref{sec:BBSAbush} \\ 
    \texttt{Spike, $2a+b \le 1$} & & poly$(n)$ \cite{Brady2016} & exp$(n)$\cite{Farhi2002} & &$O(1)$\S~\ref{sec:QAOAspike}  & $O(n)$\S~\ref{sec:BBSAspike}  \\ 
    \texttt{Spike, $2a+b > 1$} & & exp$(n)$ \cite{Brady2016} & exp$(n)$\cite{Farhi2002} & &$O(1)$\S~\ref{sec:QAOAspike}  & $O(n)$\S~\ref{sec:BBSAspike}  \\ 
    \bottomrule[1.5pt]
  \end{tabular}\\
  \caption{Performance of the four algorithms, summarized. For the two instances studied, we distinguish different parameter regimes. For the \texttt{Bush} instance, the performance of QAO depends on the choice of mixer $B_{\lambda}$ (see Eq.~\ref{eq:blambda}). For \texttt{Spike}, the QAO performance depends on spike parameters $a$ and $b$. We see that bang-bang control algorithms outperform their (quantum and classical) annealing-based counterparts for these instances. Sources for existing results are cited, and the new contributions are referenced by the relevant sections.}
\label{tab:results}
\end{table}

In addition, we study the performance of single-round QAOA (or QAOA1) on a more general class of symmetric cost functions, and give sufficient conditions under which QAOA1 can successfully find minima for these functions. These results are stated in Lemma~\ref{lem:support} and Theorem~\ref{thm:linearize}. In Sec.~\ref{sec:PMP}, we elaborate on the theoretical motivation behind choosing a bang-bang schedule and the caveats therein. 

\section{Preliminaries}
\label{sec:bg}

First, we present some notation that will be used throughout the paper. Any problem instance of size $n$ will be given as a constraint satisfaction problem on Boolean strings of length $n$.
An $n$-bit string will be expressed as a boldfaced variable, e.g. $\str{z}\in\curly{0,1}^n$, in analogy with vector quantities.
Variables denoting bits of a string will be expressed in normal font (e.g. the $i$-th bit of $\str z$ is $z_i$).
Similarly, the Hamming weight of a string, which is defined as the (integer) $1$-norm of the bit string, or the number of 1's in a bit string,
\begin{equation}
  \label{eq:HWtDef}
  |\str{z}| := \suml{i=1}{n}{z_i}
\end{equation}
will also be represented by non-bold letters such as $w, v$ to indicate that it is a scalar quantity like the value of a bit.

We will be interested in expressing states by labels such as a string variable $\str{z}$, or scalar variables $w, z$, etc. In either case, the convention will be to use $2$-normalized kets $\ket{\cdot}$, or $1$-normalized vectors, for which we will use the notation $\cket{\cdot}$. In particular, a state labeled by Hamming weight $w$ will denote the equal superposition over all bit strings with that Hamming weight,
\begin{equation}
\ket{w} := \frac{1}{\sqrt{\binom{n}{w}}}\suml{|\str z|=w}{}{\ket{\str z}},\ \ \ \cket{w} := \frac{1}{\binom{n}{w}}\suml{|\str z|=w}{}{\cket{\str z}}
\end{equation}
Problem instances are given as a cost function on bit strings,
\begin{align}
  c:&\curly{0,1}^n\rightarrow \mbb{Z}\\
  c(\str z) &= \text{cost of bit string } \str z.
  \label{eq:cost}
\end{align}
There is a natural Hamiltonian operator $C$ (and corresponding unitary $\mc C$) associated with this function that is diagonal in the computational basis, with eigenvalue $c(\str{z})$ for every corresponding eigenvector $\str{z}\in\curly{0,1}^n$. Explicitly,
\begin{equation}
  \label{eq:PhaseOpDefn}
  C := \suml{\str{z}\in\curly{0,1}^n}{}{c(\str{z})\ket{\str{z}}\bra{\str{z}}}, \ \ \ \mc C(\gamma) := e^{-i \gamma C}
\end{equation}
%Depending upon context, this is either called the phase operator, cost Hamiltonian, problem Hamiltonian, or the Hamiltonian oracle.
Classical $n$-bit strings are naturally representable as vertices of an $n$-dimensional hypercube graph. This is often the representation of choice, as walks on the hypercube are generated by sequences of bit flips on the string, which correspond to the 1-local quantum operator 
\begin{equation}
  \label{eq:MixerDefn}
  B := -\suml{i=1}{n}{X_i}, \ \ \ \mc B(\beta) := e^{-i \beta B}
\end{equation}
where $X_i \equiv \underbrace{\id\otimes \cdots\otimes\id}_{i-1}\otimes X\otimes\underbrace{\id\otimes \cdots\otimes \id}_{n-i}$, and $X = \begin{pmatrix} 0&1\\1&0\end{pmatrix}$ is the Pauli-$X$ operator. Unitary evolutions of a quantum state under $B, C$ achieve amplitude mixing and coherent, cost-dependent phase rotations, respectively. Canonically, both QAO and QAOA (see Table~\ref{tab:algs} for full names) use Hamiltonians of the form $B$ and $C$. However, as discussed towards the end of Sec.~\ref{sec:SAQAOperf}, other choices can affect the performance on a given instance.  

\begin{table}[H]
  \centering
  \begin{tabular}[h]{ccccc}
    \toprule[1.5pt]
    \head{Abbreviation} & \phantom{abc}& \head{Name of Algorithm} & \phantom{abc} & \head{Reference}\\
    \midrule[1.5pt]
    QAO & & Quantum adiabatic optimization (algorithm) & & \cite{Farhi2000} \\ 
    SA & & Simulated annealing & & \cite{Farhi2002}  \\ 
    QAOA & & Quantum approximate optimization algorithm & & \cite{Farhi2014} \\ 
    BBSA & & Bang-bang simulated annealing & & \S~\ref{sec:BBSA}\\
    \bottomrule[1.5pt]
  \end{tabular}\\
  \caption{Table of abbreviations for the algorithms studied in this paper. The last algorithm, BBSA, is introduced in this paper.}
  \label{tab:algs}
\end{table}

Now, we will describe the candidate algorithms listed in Table~\ref{tab:algs}. It will become evident that these algorithms can all be expressed in the control framework given in Appendix~\ref{sec:control}. This connection is important, as it allows us to borrow existing results from optimal control theory to the setting of heuristic optimization. 

\section{Annealing-based algorithms}
\label{sec:algA}

\subsection{Simulated annealing}
\label{sec:SA}

Simulated Annealing (SA) is a family of classical heuristic optimization algorithms that seek to minimize a potential via the evolution of a classical probability distribution under a simulated cooling process. The dynamics of the distribution are governed by two competing influences:
\begin{itemize}
\item Descent with respect to the cost function $c(\str{z})$.
\item Thermal fluctuations that kick the walker in a random uphill direction with Boltzmann \\ probability, defined according to a controlled temperature parameter $\tau$.
\end{itemize}

In practice, the above dynamics may be achieved via the following random walk:
\begin{enumerate}
\item Initialize the walker at location $\str r_1$.
\item Run a $p$-round annealing schedule, where the $i$-th round is given by $t_i\in \mbb{Z}_{\ge 0}$ time steps and a temperature parameter $\tau_i\in\mbb R_{\ge 0}$. For index $i \in [p]$, run:
  \begin{enumerate}
  \item For $t_i$ iterations, repeat:
    \begin{enumerate}
    \item Pick direction $\str{e}$ uniformly at random from available local unit displacement vectors. 
    \item Let $\delta_{\str e}:=c(\str r_i \oplus \str e) - c(\str r_i)$ be the cost increase in moving walker from current position $\str r_i$ to new position $\str r_{i+1}=\str r_i\oplus\str{e}$.
    \item If $\delta_{\str e} \le 0$, move to new location with certainty. Otherwise, move with Boltzmann probability $e^{-\delta_{\str e} /\tau_i }$, where $\tau_i$ is the current temperature in the schedule. In other words,
      \begin{equation}
        \text{Pr}(\str r_i\rightarrow \str r_i\oplus\str{e}) = \min\curly{1,e^{-\delta_{\str e}/\tau_i }}.
      \end{equation}
    \end{enumerate}
  \end{enumerate}
\item Repeat steps 1-2 several times, and report the minimal sampled configuration $\str z^*$ and the corresponding cost $c(\str z^*)$.
\end{enumerate}

The temperature schedule $\str{\tau} = \paren{\tau_1,\tau_2,\ldots,\tau_p}$ must be optimized in order to achieve a final distribution that is well-supported on low-energy states (including the global minima, ideally). In practice, one applies a finite ``cooling'' schedule in which the elements of $\str \tau$ descend from $\infty$ to $0$. At each temperature $\tau_i$, the time steps $t_i$ may be seen as relaxation time steps, where the walker distribution equilibrates under thermal exchange with the simulated bath at temperature $\tau_i$. In the limit of infinitely slow, monotonically decreasing temperature schedules that satisfy certain additional conditions arising from deep local minima in the problem instance, simulated annealing always converges to the lowest-cost configuration \cite{mitra1986,geman1984,hajek1988,gidas1985}. However, finite-time schedules and a finite relaxation time per temperature step can undo the theoretical guarantee. 

The position update of the walkers in the above scheme is implemented via the Metropolis-Hastings rule, where uphill motions are suppressed with Boltzmann probability. This implies that steeper climbs quickly become exponentially unlikely, resulting in an effective trapping of walkers in basins of depth $\sim \tau_i$. Within these basins, sufficiently high relaxation times allow the walkers to find deep minima. Intuitively, the walkers are allowed to climb barriers ``just high enough'' so as to settle into progressively deeper minima, as $\tau_i$ decreases through the course of the algorithm.

The above process may be seen as a discretization of an approximately equivalent, continuous-time Markov process. In the parlance of control introduced in Appendix~\ref{sec:control}, the dynamics of the walker distribution is generated by a stochastic operator $H\paren{\tau(t)}$ that is singly controlled by the time-dependent temperature parameter $\tau(t)$. For two neighboring positions $\str{z}, \str{z\pr}$ in the space with a mutual displacement unit vector $\str{e} = \str{z\pr} - \str{z}$, the corresponding matrix element may be written as
\begin{equation}
  H(\tau)_{\langle \str{z}\str{z\pr}\rangle} =
  \begin{cases}
    \mc{Z(\tau, \str{z})}, & \text{ if }\ \str{z}=\str{z\pr}\  \\
    1, & \text{ if }\ \delta_{\str e} \le 0\\
    e^{-\delta_{\str e} c/\tau},& \text{ if }\ \delta_{\str e} > 0
  \end{cases}
\end{equation}

where the diagonal term term $\mc Z(\tau, \str{z})$ is the negative column sum of the $\str{z}$ column of $H(\tau)$, a condition which ensures stochasticity of the Markov process. Then, the continuous-time dynamics of the probability vector are given by the differential equation
\begin{equation}
\dot{P}(t) = -H(\tau(t))P(t)
\end{equation}

Under a discretization of the above into small time slices $\Delta t_i$ (such that $||H(\tau_i)||\Delta t_i < 1$), and approximating the temperature schedule as a piecewise constant function, we may rewrite the continuous process as a Markov chain where the dynamics at the $i$-th slice are given by the stochastic matrix $\id - H(\tau_i)\Delta t_i$. This corresponds to the $i$-th step of the discrete random walk.

The infinite-temperature and zero-temperature limits of $H$ are important special cases. At $\tau=\infty$, walkers choose random directions and walk with certainty, independent of the potential. This corresponds to the case of \emph{diffusion}. On the other hand, at $\tau=0$, walkers walk in a randomly chosen direction if and only if the resulting cost is no greater than the current cost. This is what we may call \emph{randomized gradient descent}. We will denote these operators by $D, G$ respectively and give their form below:
\begin{align}
  \label{eq:DG}
  D_{\langle \str{z}\str{z\pr}\rangle} &:= H(0)_{\langle \str{z}\str{z\pr}\rangle} =
  \begin{cases}
    -n(\str z), & \text{ if }\ \str{z}=\str{z\pr}\  \\
    1, & \text{ if }\ \delta_{\str e} \le 0\\
    1,& \text{ if }\ \delta_{\str e}> 0
  \end{cases}\\
  G_{\langle \str{z}\str{z\pr}\rangle} &:= H(\infty)_{\langle \str{z}\str{z\pr}\rangle} =
  \begin{cases}
    -n_<(\str z), & \text{ if }\ \str{z}=\str{z\pr}\  \\
    1, & \text{ if }\ \delta_{\str e} \le 0\\
    0,& \text{ if }\ \delta_{\str e} > 0
  \end{cases}  
\end{align}
where $n(\str z)$ is the number of neighbors, and $n_{<}(\str z)$ the number of ``downhill'' neighbors, of $\str z$. (Note: For all bit strings $\str z$, $n(\str z) = n$ on the usual $n$-dimensional hypercube.)

%Finally, the algorithm itself may be expressed as the evolution of a probability vector $P_0$ under the time-varying $H(\tau)$ matrix,
%\begin{equation}
%  \label{eq:SAcontinuous}
%  \cket{P_T} = \prodl{i=1}{p}{e^{-H(\tau_i)\Delta t_i}}\cket{P_0}
%\end{equation}
\subsection{SA with linear update}
\label{sec:LUSA}

Under Metropolis-Hastings Monte Carlo, we see that the dynamics evolve under $H(\tau)$, which is an operator controlled by the temperature schedule $\str \tau$. The obvious bang-bang analogue to this is to alternate between periods of zero- and infinite-temperature Metropolis moves, which is the algorithm introduced in Sec.~\ref{sec:BBSA}. However, to argue that bang-bang control is optimal using the optimal control framework (as in Sec.~\ref{sec:PMP}), we must first ensure that the dynamics are linear in the controls. In this section, we present a linearized variant of SA, so that within algorithms of this class, it will be the case that bang-bang control is optimal as a consequence of the Pontryagin Minimum Principle. 

Suppose that instead of Metropolis-Hastings probability $\min\curly{1, e^{-\delta_{\str e}/\tau}}$, we use a probability $u\Theta(\delta_{\str e})$, where $\Theta (\cdot)$ is the Heaviside step function, and $u \in [0,1]$ is a control parameter. That is, 
\begin{equation}
  \label{eq:LUSAupdate}
  \text{Pr} (\str z \rightarrow \str {z}\oplus\str{e}) =
  \begin{cases}
    1, & \text{ if }\ \delta_{\str e} \le 0\\
    u, & \text{ if }\ \delta_{\str e} > 0
  \end{cases}
\end{equation}
This rule is qualitatively different from Metropolis-Hastings, since it attaches importance not to the exact energy difference between neighboring states, but only to its sign. Furthermore, the update rule is not guaranteed to satisfy physical prerequisites such as detailed balance that guarantee the convergence of the limiting distribution. However, it is a valid update rule, and we will call SA equipped with these dynamics linear update SA.

Importantly, linear update SA is expressible in the linear control framework. It is possible to write the Markov matrix $H(u)$ corresponding to the continuous version of Eq.~\ref{eq:LUSAupdate} as a sum of the diffusion matrix D and the randomized gradient descent operator $G$,
\begin{equation}
  \label{eq:LUSAlinearcontrol}
  H(u) = uD + (1-u)G
\end{equation}
Finally, we see that, $H(u = 0)  = D$ and $H(u = 0) = G$, thus reproducing the operators appearing in standard SA in the limit of infinite and zero temperature (i.e. $u = 0,1$), which are the relevant parameter values under bang-bang control.

\subsection{QAO}
\label{sec:QA}
The adiabatic algorithm, proposed in 2000 by Farhi, Goldstone and Gutmann \cite{Farhi2000}, is a (quantum) heuristic combinatorial optimization algorithm based upon the adiabatic theorem from quantum mechanics. The adiabatic theorem, loosely stated, says that a system evolving under a time-varying Hamiltonian, when initialized in a ground state, stays in the instantaneous ground state as the Hamiltonian is varied \emph{slowly} in time. The recipe to turn this statement into an algorithm for finding global minima is as follows:
\begin{enumerate}
\item Initialize the system in an easily preparable ground state of a Hamiltonian $B$. 
\item Read the problem instance (cost function $c(\str z)$), and map it to an equivalent Hamiltonian $C$, as in Eq.~\ref{eq:PhaseOpDefn}.
\item Implement Schr\"{o}dinger evolution of the state over the time interval $[0,T]$ under a controlled Hamiltonian $H(s) = u_1(s)B + u_2(s)C$, where $s=t/T$ is the scaled time parameter, and $u_1,u_2$ are functions of $s$ that describe the \emph{annealing schedule}. The schedule satisfies $u_1(0)=1 -u_2(0)= 1$, and $u_1(1)=1-u_2(1)=0$.
\item Measure the resulting state in the computational basis.
\end{enumerate}

Under adiabaticity (i.e. when the schedule varies slowly in $s$), the above algorithm evolves the initial state from the ground state of $B$ to that of $C$, which is a state that encodes the solution to optimization problem. In particular, the algorithm succeeds if the rate is slower than inverse polynomial in the first spectral gap $\lambda(s)$ (i.e. the energy difference between the ground state and the first excited state) at all times. Typically, this yields a condition on the true runtime $T$ \cite{jansen2007, elgart2012}:
\begin{equation}
  \label{eq:QAruntime}
  T \gtrsim O\paren{\frac{1}{\lambda^2}}
\end{equation}
where $\lambda = \min_s\lambda_s$. Therefore, the guarantee of success of an adiabatic protocol lies in knowing that the minimum gap $\lambda$ does not scale super-polynomially with $n$. However, it should be noted that this does not rule out good empirical performance. In fact, by cleverly varying speed as a function of the instantaneous spectral gap, important speedups such as the Grover speedup \cite{roland2002}, and the exponential speedup for glued trees, \cite{somma2012} can be recovered.

Like Linear Update SA, the QAO Hamiltonian
\begin{equation}
  H(u_1,u_2) = u_1B + u_2C
  \label{eq:qlincontrol}
\end{equation}
fits into the linear control framework \cite{mcclean2016}. In fact, we may simplify the above Hamiltonian to a singly-controlled Hamiltonian as follows. In practical applications of QAO, there is a maximum magnitude threshold (say $J$) for the controls, given by hardware constraints. We assume that this cutoff does not scale with the input size of the instance. Assume also that the lower cutoff for both $u_1$ and $u_2$ is $0$. In other words, $u_1, u_2 \in [0,J]$. These design constraints give us a restricted version of QAO where the controls are non-negative and bounded. This restriction is applied simply to state our algorithms within a uniform, linear control framework. Adiabatic algorithms for the instances studied in later sections fit within this framework. 

Then, observe that when $u_1 + u_2 > 0$, we can rescale the controls by factor $u_1(s) + u_2(s)$, giving us the following mapping of the time variable and the controls:
\begin{align}
  \frac{\text{d}s}{\text{d}t} &\mapsto \frac{\text{d}s}{\text{d}t}\cdot\paren{u_1(s) + u_2(s)}\\
  \paren{u_1(s), u_2(s)} &\mapsto \paren{u := \frac{u_1(s)}{u_1(s) + u_2(s)}, 1-u} 
\end{align}

Under this mapping, the time parameter is rescaled by a factor of at most $2J$ (corresponding to a constant slowdown), while the parametric Hamiltonian now looks like
\begin{equation}
  \label{eq:QAOlinearcontrol}
  H(u_1, u_2) \mapsto H(u) = uB + (1-u)C
\end{equation}

When $u_1(s) = u_2(s) = 0$, which is the only case not covered by the above mapping, we see that the dynamics ``switch off'' completely. This feature is useful only when the total time $T$ is greater than the time necessary to complete the algorithm. However, if we study protocols as a function of the time horizon $T$, this feature becomes unnecessary, and we may safely ignore it.

Therefore, we have successfully mapped QAO to a linear, single control framework with only a constant overhead in the run time. From now on, we assume that QAO possesses the form given in Eq.~\ref{eq:QAOlinearcontrol}.

\section{Bang-bang algorithms}
\label{sec:algB}
In parallel with the developments in annealing-based methods, extensive studies have been conducted into the problem of optimal control of quantum dynamics (see \cite{chakrabarti2007}), particularly in the context of many-body ground state preparation, e.g \cite{rahmani2011}. It is often found to be that case that, contrary to a quasistatic schedule, a rapidly switching, bang-bang schedule could be engineered to prepare states quickly.

In combinatorial optimization, an alternative framework based on circuits with variable parameters has been investigated, and has recently gained interest with the introduction of the Quantum Approximate Optimization Algorithm (QAOA), \cite{Farhi2014}. This is in fact an example of bang-bang control, as observed in \cite{mcclean2016}. The related problem of ground state preparation has also been approached using Variational Quantum Eigensolver (VQE) ans\"{a}tze \cite{peruzzo2014} that bear close resemblance to QAOA in their setup. The recent work by Hadfield et al. \cite{hadfield2017} has proposed a relabeling of the acronym QAOA to the `Quantum Alternating Operator Ansatz' to capture this generality. In this manner, a new path that explores classical design strategies of quantum algorithms, also known as a hybrid approach, has been paved. 

In the coming sections, we will formally introduce QAOA, as well as a new, classical bang-bang version of SA which we call bang-bang SA, or BBSA. Then in Sec.~\ref{sec:PMP}, we will elaborate on the theoretical motivation behind choosing the bang-bang approach.  

\subsection{Bang-bang simulated annealing (BBSA)}
\label{sec:BBSA}

BBSA is the restriction of linear update simulated annealing (see Sec.~\ref{sec:LUSA}) to bang-bang schedules. In other words, this is an algorithm that alternately applies diffusion and randomized gradient descent to the state. An instance of this algorithm may then be specified by the number of rounds $p$ (where in each round we apply the two operators in succession), and the corresponding evolution times for each round.

Observe that Metropolis-Hastings SA, when restricted to $\tau = 0, \infty$, reduces to bang-bang SA.

\subsection{QAOA}
\label{sec:QAOA}

The Quantum Approximate Optimization Algorithm (QAOA) was introduced by Farhi \emph{et al.} in 2014, \cite{Farhi2014}, as an alternative ansatz to the QAO. We note (as is done in \cite{mcclean2016}) that, like QAO, QAOA is a restriction of the linearly controlled Hamiltonian to the case of bang-bang control, i.e., where we only allow $u = 0, 1$ at any given time.

Restricted in this way, a QAOA protocol effectively implements a series of alternating Hamiltonian evolutions under the mixing operator $B$, and the cost operator $C$. Therefore, for a total of $p$ rounds of alternating evolution with evolution angles $\vec{\beta}:= \paren{\beta_1,\ldots, \beta_p}, \vec{\gamma}:= \paren{\gamma_1,\ldots, \gamma_p}$ for $B$ and $C$ respectively, the final state prepared by QAOA may be expressed as
\begin{equation}
  \ket{\vec{\beta}, \vec{\gamma}} = \brac{\prodl{i=1}{p}{\mc B(\beta_i)\mc C(\gamma_i)}}\ket{\psi_0}
\end{equation}
where we used the parameterized operators from Eq. \ref{eq:PhaseOpDefn}, \ref{eq:MixerDefn}, and, as in the case of QAO, the initial state $\ket{\psi_0}$ is an easily preparable state such as the equal superposition of bitstrings, $\ket{+^{\otimes n}}$.

QAOA with a fixed number of rounds $p$, also written as QAOA$p$, is a scheme for preparing one of a family of trial states of the form $\ket{\vec{\beta}, \vec{\gamma}}$. With the angles as search parameters, a figure of merit such as the energy expectation of the cost operator $E(\vec \beta, \vec \gamma) = \bra{\vec{\beta}, \vec{\gamma}} C \ket{\vec{\beta}, \vec{\gamma}}$ is approximately minimized with the aid of classical outer loop optimization. 

\section{Conditions for optimality of bang-bang control}
\label{sec:PMP}

Now, we will elaborate on the theoretical motivation for choosing a bang-bang approach to optimization algorithms, expanding on the observations made in \cite{mcclean2016, Yang2017}. The Pontryagin Minimum Principle (PMP) from optimal control theory \cite{pontryagin} provides key insight into the nature of optimal schedules for heuristic optimization algorithms expressible in the control framework. As discussed in Appendix~\ref{sec:control}, PMP gives necessary conditions on the control in the form of a minimization of the control Hamiltonian, which is a classical functional of the state amplitudes and corresponding conjugate ``momenta'', and depends on the control parameters as well. 

When the control Hamiltonian $\mc{H}$ is linear in the control vector $\str u$, the minimization condition Eq. \ref{eq:pmp} implies that the optimal control is \emph{extremal}, in the sense that the control only takes values on the boundary of the feasible control set at any given time. When the control parameters are individually constrained to lie in a certain interval, $u_i\in [a_i,b_i]$, then we say that the optimal protocol is bang-bang, i.e. $u_i(t) = a_i\ \text{ or}\ b_i$. Thus, the individual controls switch between their extremal values through the course of the protocol. 
While the heuristic algorithms QAO, QAOA and SA with linear update satisfy the condition of linear control, one should exercise caution when stating the optimality of bang-bang control within these frameworks. We note a few important caveats here:
\begin{enumerate}
\item PMP simply gives a necessary condition for optimality, it does not provide the optimal protocol. A different control theory tool, the Hamilton-Jacobi-Bellman equation, does provide a way to find the optimal protocol via dynamic programming. 
\item There may be an arbitrary number of switches in the optimal bang-bang protocol. In fact, some problems exhibit the so-called \emph{Fuller phenomenon}, in which the optimal control sequence has an infinite number of bangs, and is therefore rendered infeasible. 
\item The control Hamiltonian may become \emph{singular} at any point during the protocol. A singular interval is one in which the first derivative of $\mc H$ with respect to $u$ vanishes. In these intervals, the optimal control is not necessarily bang-bang. The presence of generic singular intervals has already been observed before in the dynamics of spin systems (see, e.g. \cite{lapert2011, bonnard2012}). Therefore, in order to guarantee that the optimal control is bang-bang at all times, one must first show that there are no singular intervals during the protocol. 
\item The original PMP is stated and proved for dynamics over Euclidean vector spaces over $\mbb{R}$. However, in quantum optimization the amplitudes take values in $\mbb{C}$, and the Hilbert space is a complex projective vector space with a non-Euclidean geometry. The generalization must be made with caution.
\end{enumerate}
Despite these caveats, PMP does provide theoretical motivation for using bang-bang control as a design principle for heuristic optimization algorithms. In the following sections, we exhibit examples where bang-bang control exponentially outperforms conventional SA and QAO.

\section{The problem instances}
\label{sec:probInsts}

Now, we describe the problem instances that will be used as benchmarks for our algorithms. The two following instances have appeared in the context of comparisons between quantum and classical heuristic optimization algorithms, usually to show the inability of the classical algorithm to escape a local minimum and find the true, global minimum \cite{Farhi2002, Brady2016, Kong2017}. This is often interpreted as evidence of a quantum advantage, such as the ability to tunnel through barriers. In keeping with this tradition, we will select these as our benchmarking instances, and look for general features in the performance of our candidate algorithms. 

\begin{figure}[H]
  \centering
  \includegraphics[scale=0.28, clip=true, trim= 60pt 0pt 70pt 0]{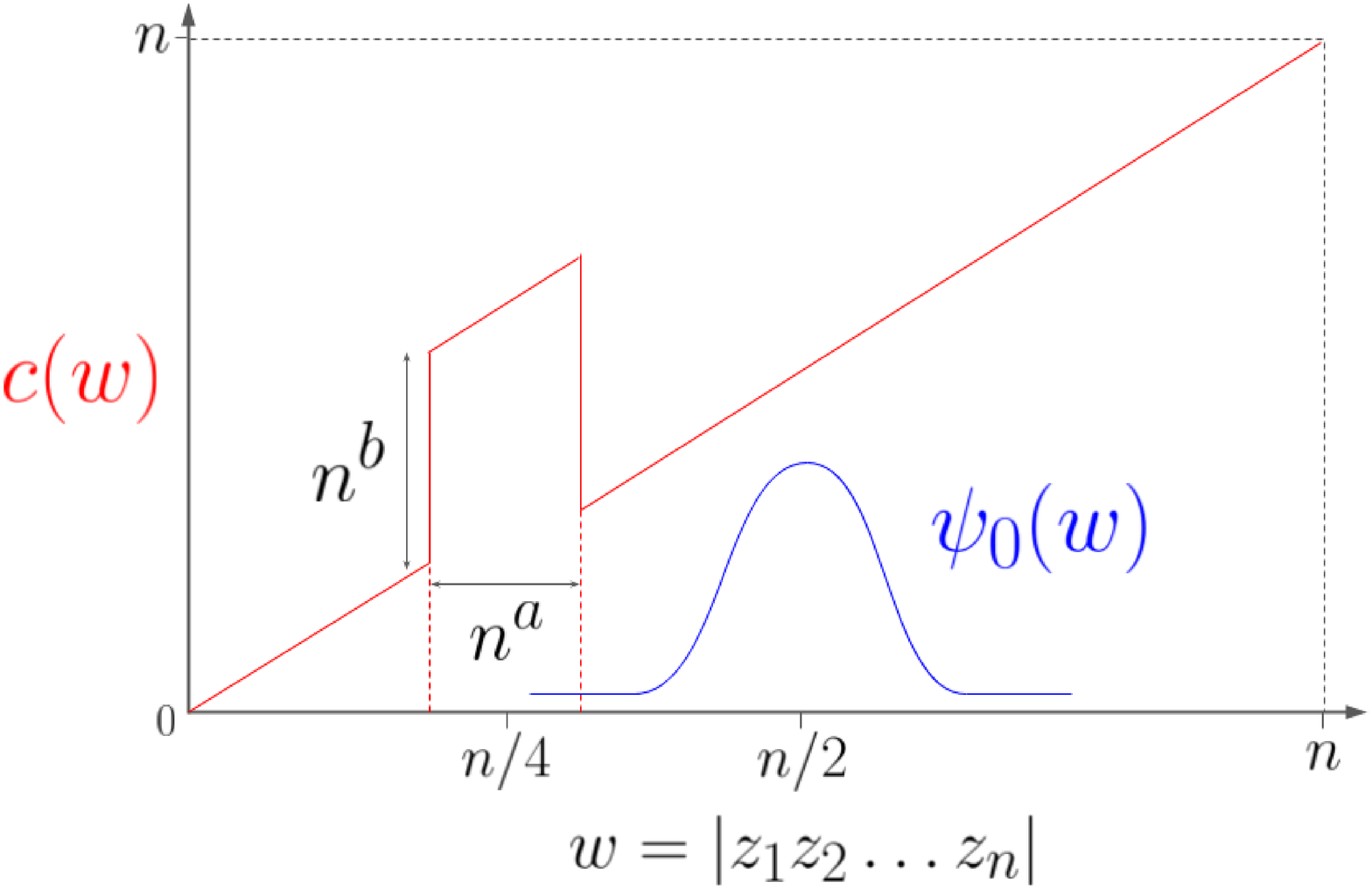}\includegraphics[scale=0.28, clip=true, trim= 60pt 0pt 70pt 0]{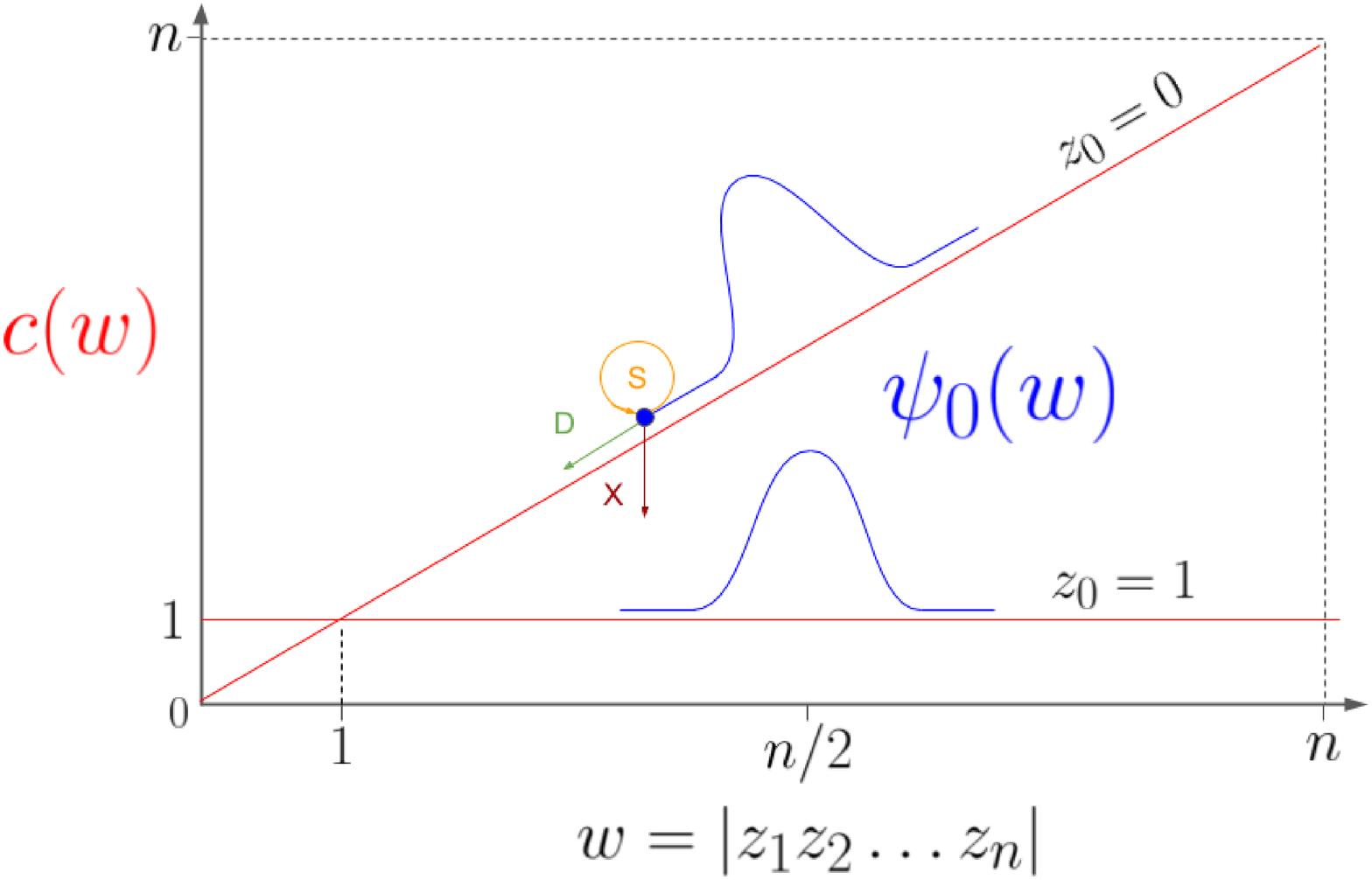}
  \caption{Schematic energy landscapes of the two instances, \texttt{Spike} (left) and \texttt{Bush} (right). In each diagram, the blue curve indicates the distribution of the initial state, the equal superposition over all bit strings.}
  \label{fig:instances}
\end{figure}

\subsection{Bush of implications}
\label{sec:instances}
The bush of implications or \texttt{Bush} is an instance first crafted in \cite{Farhi2002} in order to demonstrate the failure of SA where QAO succeeds, with an exponential separation between the two. In \texttt{Bush}, the potential is not fully symmetric under permutation of bits. Instead, the first bit (the ``central'' bit, indexed by 0) determines the potential acting on the Hamming weight of the remaining $n$ ``peripheral'' bits. Specifically,

\begin{equation}
c(\str z = z_0z_1\ldots z_n) = z_0 + \suml{i=1}{n}{z_i\paren{1-z_0}} \\ = z_0 + w\paren{1-z_0}\label{eq:pot}
\end{equation}
where $w = |z_1\ldots z_n|$. So, the potential is constant and equal to 1 when $z_0=1$, and a Hamming ramp, $r(w)=w$ when $z_0=0$, as shown in Fig.~\ref{fig:instances}. Note that we adopted a bit-flipped definition of $c$ as compared to the original in \cite{Farhi2002}. The reason is simply notational convenience. The energy landscape of the bush of implications can be viewed as the number of clauses violated in a constraint satisfaction problem, where each clause takes the form $\neg z_0\implies \neg z_j$ for $j>0$, which lends the instance its name.

\subsection{Hamming ramp with spike}
Next, we present a second family of Hamming-symmetric potentials studied first in \cite{Farhi2002, Reichardt2004}, the Hamming ramp with a spike. In the general form more recently studied in \cite{Brady2016,Crosson2016, Kong2017}, this potential is given by a ramp $r(w) = w$, plus a rectangular ``spike'' function $s(w)$ centered at $w=n/4$ with width $O(n^{a})$ and height $O(n^{b})$, for two exponents $a, b \in [0,1]$.
\begin{align}
  \label{eq:spike}
  \text{Ramp: }\ r(w) = w,\ &\text{Spike: }\ s(w) = \begin{cases} n^b,\text{ if } w\in [\frac{n}{4}-\frac{n^a}{2}, \frac{n}{4}+\frac{n^a}{2}]\\0,\text{ otherwise.}\end{cases}\\
  \text{Full Potential: }\ c(w) &= r(w) + s(w)
\end{align}
We will use this form for the \texttt{Spike} family of instances. 

\section{Performance}
\label{sec:performance}
Now, we will state the performance of the algorithms from Sec.~\ref{sec:algA}, \ref{sec:algB} on the instances defined in Sec.~\ref{sec:instances}, deriving or using existing results as appropriate. We will find that in both the classical (SA vs. BBSA) and the quantum (QAO vs. QAOA) settings, there exist parameter regimes in which the bang-bang algorithms are exponentially faster than their quasistatic analogues. 

\subsection{SA and QAO}
\label{sec:SAQAOperf}
For both the \texttt{Bush} and \texttt{Spike} examples, Farhi et al. argue in \cite{Farhi2002} that simulated annealing gets stuck in local minima, and is exponentially unlikely to reach the global minimum in polynomial time, in the input size $n \rightarrow \infty$. Additionally, they argue for the success of QAO on these instances in certain parameter regimes.

For the \texttt{Spike} example, \cite{Reichardt2004} and \cite{Crosson2016} show that when the width and height parameters satisfy $a+b \le 1/2$, quantum annealing solves \texttt{Spike} efficiently. If, on the other hand, $2a + b > 1$, it was shown by \cite{Brady2016} that the minimum spectral gap has an exponential scaling in $n$, implying the failure of quantum annealing in this problem regime. For the \texttt{Bush} example, it was shown in \cite{Farhi2002} that the gap scaling is polynomial in $n$, thus allowing for an efficient adiabatic algorithm to solve this instance. We note that the performance depends on the choice of the initial mixing Hamiltonian $B$. In particular, out of the following family of mixers
\begin{equation}
  \label{eq:blambda}
  B_\lambda = -\lambda(n+1)X_0 - \suml{i=1}{n}{X_i}, 
\end{equation}
QAO is successful when $\lambda \ge 1$. On the other hand, when $\lambda=1/(n+1) < 1$, we recover the canonical mixing operator $B$ from Eq.~\ref{eq:MixerDefn}, and QAO is expected to take exponential time to solve \texttt{Bush}.

Despite the caveats, \texttt{Bush} and \texttt{Spike} are examples of instances where we have an exponential separation between a quantum (QAO) and classical (SA) algorithm. However, in the next section we show that a different, purely classical, bang-bang strategy matches the performance of QAO on the \texttt{Bush} and \texttt{Spike} instances by solving them in polynomial time. 
\subsection{Bang-bang simulated annealing}
\label{sec:BBSAperf}
Now, we will show that the bang-bang version of simulated annealing is able to find the ground state of both \texttt{Bush} and \texttt{Spike} in time polynomial in $n$, and therefore exponentially outperforms SA (and QAO for certain parameter regimes, see Table~\ref{tab:results}), on both instances.

\subsubsection{\texttt{Bush}}
\label{sec:BBSAbush}
We will now show that BBSA efficiently finds the minimum of \texttt{Bush} via BBSA. In fact, the protocol simply involves performing randomized gradient descent ($G$) without any switches to diffusion. First, we characterize the $G$ matrix for this instance. The natural basis for this problem is a conditional Hamming basis $\curly{\cket{z_0,w}: z_0\in[1], w\in [n]}$ parameterized by the value of the central bit $z_0$, and the weight of the peripheral string $w = |z_1\cdots z_n|$. The allowed transitions under $G$ are as given below:
\begin{align}
  \cket{0,w} &\rightarrow \cket{1, w},\ \text{for all }\ w>0.\\
  \cket{z_0,w} &\rightarrow \cket{z_0,w-1},\ \text{for all }\ z_0\in[1],\ w>0.\\
  \cket{1,0} &\rightarrow \cket{0,0}.
\end{align}
In particular, this implies that a walker at the global minimum $\cket{0,0}$ cannot leave under $G$. Consider a discrete, Markov chain Monte Carlo implementation of $G$, in which we break up the Markov evolution into $N = 1/\delta t$ steps of size $\delta t$. The stepsize $\delta t$ is an empirical parameter which will be set later, while at the moment we only assume that $\delta t \ll 1$. Then, we may write the Markov evolution as
\begin{equation}
  \label{eq:BoIdiscreteMarkov}
    \cket{P_N} = \brac{\prodl{i=1}{N}{e^{-G\delta t}}}\cket{P_0}\simeq \brac{\prodl{i=1}{N}{\paren{\id - G\delta t}}}\cket{P_0}
\end{equation}
Each step $\id - G\delta t$ above is a stochastic evolution if $\delta t$ is sufficiently small, i.e., if all entries of the matrix represent valid probabilities. The requirement that the column sum be 1 is automatically satisfied since $G$ is column-sum-zero. Then, we start with a walker sampled from the initial state $\cket{P_0}$, and, for every step $1$ to $N$, we update the walker's position based on the transition probabilities given by $\id - G\delta t$. This is given in more detail below. We will show that the above procedure transports a fraction of at least $n^{-2.503}$ of walkers to the global minimum, in number of steps $N = O\paren{\frac{1}{\delta t}\log n}$. Finally, arguing that it suffices to choose $\delta t = \Theta (n^{-1})$ gives a polynomial runtime of $\Theta (n^{3.503}\log n)$ to have a constant success probability.

In our analysis, we only keep track of the walker in the $z_0=0$ subspace, which contains the global minimum. Any walker that starts in or enters the $z_0=1$ subspace during the algorithm will be presumed dead, and we terminate its walk. This simplification is allowed, since it may only worsen the success probability obtained through this analysis. Initially, exactly half of the walkers are alive, i.e. in the subspace $z_0=0$, and concentrated in a band of width $\sim \sqrt{n}$ around $w=n/2$. For a walker at Hamming weight $w>0$, there are three possible moves (illustrated in Fig.~\ref{fig:instances}):
\begin{enumerate}
\item ($D$) Descend to weight $w-1$, with probability $w\delta t$.
\item ($S$) Stay at the same location with probability $1 - (w+1)\delta t$.
\item ($X$) ``Die'', i.e., escape to the $z_0=1$ subspace, with probability $\delta t$.
\end{enumerate}
When $w=0$, the $D$ and $X$ moves are forbidden, and the walker can only stay in place. Additionally, we denote the event of survival (i.e. $D$ or $S$) by $\bar X$. Now, we track the random walk under the stated moves. Let $\hat m$ be a random variable representing the total number of moves the walker takes to reach the global minimum, $\cket{0,0}$. If the walker dies, we say that $\hat m = \infty$. Otherwise, $\hat m$ is finite and equal to the sum of number of moves spent at each weight $w = 1,2,\ldots, n$. Defining a corresponding random variable $\hat m_w$ for the number of moves spent at each weight, we may write
\begin{equation}
  \label{eq:BoIrandomDescent}
  \hat m = \suml{w=1}{n}{\hat m_w}
\end{equation}
The expected value of $\hat m$ tells us how many moves any given walker needs to reach the global minimum under $G$. However, since we are only interested in living walkers, we will condition the expectation on the walker staying alive ($\bar X$). Then,
\begin{equation}
  \mbb{E}\paren{\hat m~|~\bar X} = \suml{w=1}{n}{\mbb{E}\paren{\hat m_w~|~\bar X}}
\end{equation}
At each weight $w$, the condition of survival limits the allowed moves to the regular expression $S^*D$. In other words, the walker stays in place for some number of moves before descending. Note that the probability of not dying in $m$ moves is $(1-\delta t)^m$. Therefore, the probability of spending $m$ total moves, conditioned on survival, is given by
\begin{align}
  \prob\paren{\hat m_w = m~|~\bar X} &= \frac{\prob\paren{S^{m-1} D}}{\prob\paren{\bar{X}^m}} = \frac{\paren{1-\paren{w+1}\delta t}^{m-1}w\delta t}{\paren{1-\delta t}^m}\\
  & = \paren{\frac{1-\paren{w+1}\delta t}{1-\delta t}}^{m-1}\cdot \frac{w\delta t}{1-\delta t} \\ &\lesssim e^{-w\delta t\paren{m-1}}\frac{w\delta t}{1-\delta t} %\simeq  w\delta t\cdot e^{-mw\delta t}
\end{align}
where the last inequality follows from a Taylor series comparison of $\paren{1-(w+1)\delta t}/{1-\delta t}$ under the assumption that $\delta t < 1$. So, the expectation value of $\hat m_w$ is
\begin{equation}
  \mbb{E}\paren{\hat m_w~|~\bar X} = \suml{m=1}{\infty}{m \cdot \prob\paren{m~|~\bar X}} \lesssim \frac{w\delta t\cdot e^{w\delta t}}{1-\delta t}\suml{m=1}{\infty}{m \cdot e^{-mw\delta t}} = \frac{w\delta t}{\paren{1-\delta t}\paren{1-e^{-w\delta t}}^2}
\end{equation}
Finally, the full expectation value is given by
\begin{equation}
  \label{eq:BoIexpecation}
  \mbb{E}\paren{\hat m~|~\bar X} \lesssim \frac{1}{1-\delta t}\suml{w=1}{n}{\frac{w\delta t}{\paren{1-e^{-w\delta t}}^2}}
\end{equation}
Next, using the variable substitution $x = w\delta t, dx = \delta t$, we may turn the above sum into an approximate integral. In fact, the integrand $x/(1-e^{-x})^2$ is monotonically decreasing, so the sum is upper bounded by
\begin{align}
  \mbb{E}\paren{\hat m~|~\bar X} &\lesssim  \frac{\delta t}{\paren{1-\delta t}\paren{1-e^{-\delta t}}^2} + \frac{1}{\delta t(1-\delta t)}\intl{\delta t}{n\delta t}{\frac{x}{\paren{1-e^{-x}}^2}dx}\\
                                       &\lesssim \frac{4}{\paren{1-\delta t}\delta t} + \frac{1}{\delta t(1-\delta t)}\intl{\delta t}{n\delta t}{\frac{4}{x}dx} = \frac{4}{\paren{1-\delta t}\delta t} + \frac{4}{\delta t(1-\delta t)}\log(n)
%  &= O\paren{\frac{4}{\delta t}\log n}
\end{align}
where we used the trick that since $x/2$ and $1-e^{-x}$ are both monotonically increasing, and $x/2 < 1-e^{-x}$ for $x=0,1$, then it follows that $x/2 < 1-e^{-x}$ for all $x\in[0,1]$. In fact, a tighter bound may be obtained by replacing $2$ by $e/(e-1)\approx 1.58$, which yields a scaling of $\mbb{E}\paren{\hat m~|~\bar X}\lesssim \frac{2.503}{\delta t}\log n$.
Finally, the expected survival probability is $\prob \paren{\bar X} \gtrsim e^{-\delta t\cdot \frac{2.503}{\delta t}\log n} = \frac{1}{n^{2.503}}$, which is polynomial in $n$. Therefore, applying this algorithm for $\frac{1}{\delta t}\log n$ with $\delta t = \Theta(n^{-1})$, yields a polynomial probability of success. Repeating for at most $n^{2.503}$ trials amplifies the success probability to a constant. So, the total time complexity is $O{n^{3.503}\log n}$, which is efficient in the input size $n$.

In Fig.~\ref{fig:BushNumerics} below, numerics of the continuous-time process (see Eq.~\ref{eq:BoIdiscreteMarkov}) confirm that the total time indeed scales as $\log n$.
\begin{figure}[H]
  \centering
  \includegraphics[scale=0.5]{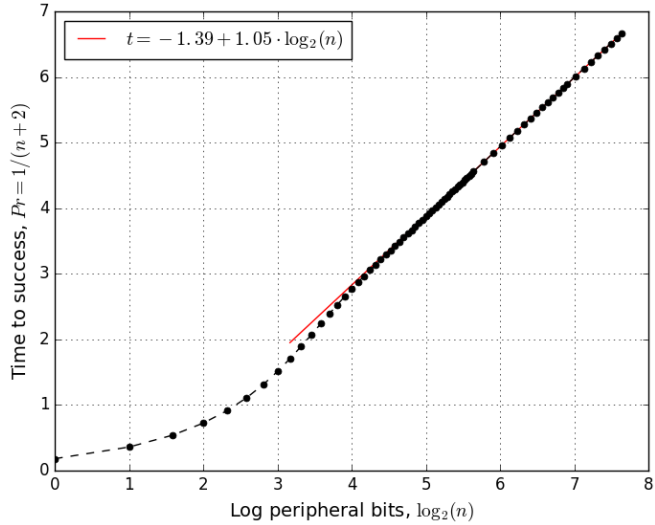}
  \caption{Plot of the input size $n$ vs. total time for success (determined by the time taken for a polynomial fraction of walkers to reach the global minimum). Note that the continuous-time process does not contain the polynomial factors; those arise from discretization into small timesteps $\delta t$ of order $\lesssim 1/n$.}
  \label{fig:BushNumerics}
\end{figure}

\subsubsection{\texttt{Spike}}
\label{sec:BBSAspike}
In the previous section, we showed that \texttt{Bush} is a problem instance where classical bang-bang algorithm (BBSA) can outperform a classical quasistatic algorithm (SA) exponentially. While this suffices to show the polynomial inequivalence of SA and BBSA, it is nonetheless interesting to explore further examples where this is the case. The \texttt{Spike} problem, as presented in \ref{eq:spike}, is the second instance where BBSA can exponentially outperform SA and QAO. Since the separation is sensitive to details such as the shape of the spike, we refer the reader to Appendix~\ref{sec:BBSAspikeApp} for further discussion.

\subsection{QAOA} 
\label{sec:QAOAperf}
Lastly, we will show that one round of QAOA (or QAOA1) efficiently finds the minimum of the instances \texttt{Bush} and \texttt{Spike}. In fact, as discussed later in this section, QAOA1 solves a more general class of symmetric instances that includes the \texttt{Spike} (and with some more analysis, the \texttt{Bush}) example. This is one of the main results of the paper, given in Theorem~\ref{thm:linearize}.

\subsubsection{\texttt{Spike}}
\label{sec:QAOAspike}
One of the key features of this instance is that the spike has exponentially small overlap with the initial state $\ket{+}^{\otimes n}$. Intuitively, this implies that the state does not ``see'' the spike, and should therefore behave as if evolving under a pure Hamming ramp. We state this as the following lemma:

\begin{lemma}
\label{lem:support} Let c(w) be a Hamming-symmetric cost function on bitstrings of size $n$, and let $p(n) \in [0,1]$ be a problem size-dependent probability. Suppose $c(w) = r(w) + s(w)$, where $r, s$ are two functions satisfying the following:
\begin{enumerate}
\item $\min_{w}c(w) = \min_{w}r(w)$.
\item There exist angles $\beta, \gamma$ such that QAOA1 with schedule $(\beta, \gamma)$ minimizes $r(w)$ with probability at least $p(n)$.
\item If the initial state is $\ket{\psi_0} = \sum_w A_w\ket{w}$, then $s(w)$ overlaps weakly with $\ket{\psi_0}$ in the sense that
  $$
  \suml{w=1}{n}{4|A_w|^2\sin^2\paren{\frac{\gamma s(w)}{2}}} \le  o(p(n))
  $$
\end{enumerate}
Then, QAOA1 with schedule $(\beta, \gamma)$ minimizes $c(w)$ with probability at least $p(n) - o(p(n))$. 
\end{lemma}
For the \texttt{Spike} instance, we decompose the cost into a ramp term and a spike, $c(w) = r(w) + s(w)$. First, we compute the success probability of QAOA1 on only the ramp term $r(w)$. This potential may be written as 
\begin{equation}
  \label{eq:SpikeTensor}
  R = \suml{w=0}{n}{w\ket{w}\bra{w}} = \suml{i=1}{n}{\frac{\id - Z_i}{2}} = \frac{n}{2}\id - \frac{1}{2}\suml{i=1}{n}{Z_i}
\end{equation}
which is a 1-local operator on qubits, just like $B$. It can be seen that the protocol simply applies a rotation from the $\ket{+}$ state to the $\ket{0}$ state on each qubit via a $Z/2$ rotation followed by an $X$ rotation, and succeeds with probability 1. The angles can be read off from the Bloch sphere: $\gamma = 2\cdot\pi/4 = \pi/2$, and $\beta = \pi/4$.

Then, it follows from Lemma~\ref{lem:support} that the effect of the spike $s(w)$ under QAOA1 is negligible if $\sum_w4\sin^2(\gamma s(w)/2)|A_w|^2$ is small, where $A_w$ are amplitudes of the initial state in the symmetric basis. But this sum may be bounded as
\begin{align}
  \label{eq:SpikeBound}
  \suml{w=0}{n}{4\sin^2(\gamma s(w)/2)|A_w|^2} &= \frac{1}{2^{n-2}}\suml{w=n/4-n^a/2}{n/4+n^a/2}{\sin^2(\gamma n^b/2)\binom n w} \\
  &\le \frac{1}{2^{n-2}}\suml{w=n/4-n^a/2}{n/4+n^a/2}{\binom n w} = 4\suml{w=n/4-n^a/2}{n/4+n^a/2} B(w;n,1/2)
\end{align}
where $B(w;n,1/2)$ is a binomial term corresponding to the probability of $n$ tosses of a fair coin returning exactly $w$ heads. Now, we may use known bounds on tail distributions such as Hoeffding's inequality, and we finally have
\begin{equation}
   \suml{w=0}{n}{4\sin^2(\gamma s(w)/2)|A_w|^2} = 4\suml{w=n/4-n^a/2}{n/4+n^a/2} B(w;n,1/2) = o(1) \text{ when } a < 1
\end{equation}
Then, applying Lemma~\ref{lem:support}, we conclude that, for a spike with $a\in [0,1)$ and arbitrary $b$, QAOA1 with angles $(\pi/4,\pi/2)$ finds the global minimum with probability polynomially close to 1. 

This QAOA1 protocol is asymptotically successful for any $\paren{a,b}$ chosen from the set $[0,1)\times \mbb R$. In practice, finite $n$ instances will show effects of the finite overlap of the initial state with the spike at $a$ close to 1. But even in this regime, the barrier height is essentially irrelevant, since it appears in the argument of a sinusoid and may only affect the bounds in Eq.~\ref{eq:SpikeBound} by a constant.

\subsubsection{\texttt{Bush}}
\label{sec:QAOAbush}
The \texttt{Bush} instance is a quasi-symmetric potential, since it depends on the value of the central bit. In the $z_0=1$ sector, the potential is a constant, while in the $z_0=0$ sector, it is a ramp. So, in analogy with Eq.~\ref{eq:SpikeTensor}
\begin{equation}
  \label{eq:BushTensor}
C = \ket{1}\bra{1}\otimes\id + \ket{0}\bra{0}\otimes\paren{\frac{n}{2}\id - \frac{1}{2}\suml{i=1}{n}{Z_i}}
\end{equation}
For ease of analysis, separate the mixing operator into the mutually commuting peripheral terms and central term:
$$
\mc{B}(\beta) = e^{-i\beta B} = \paren{\cos\beta \id_0 - i\sin\beta X_0}\prodl{i=1}{n}{\paren{\cos\beta \id_i - i\sin\beta X_i}} \equiv \mc{B}_0\mc{B}_i
$$ 
As before, the QAOA protocol implements one $Z$ rotation (operator $\mc C(\gamma) = e^{-i\gamma C}$) followed by an $X$ rotation (operator $\mc B(\beta)$). Since the \texttt{Bush} potential contains a ramp in the relevant sector, we will try the protocol used for the \texttt{Spike} instance, $\beta=\pi/4,\gamma=\pi/2$.

The $Z$-rotation transforms the initial state (on the peripheral bits) into the $+Y$ eigenstate, $\ket{+}^{\otimes n} \rightarrow \frac{1}{\sqrt{2^n}}\paren{\ket{0}+i\ket{1}}^{\otimes n}$. So, the full state transforms as
\begin{equation*}
\frac{1}{\sqrt{2}}\ket{1}\otimes\ket{+}^{\otimes n} + \frac{1}{\sqrt 2}\ket{0}\otimes\ket{+}^{\otimes n}\ \underset{\mc C(\pi/2)}{\xrightarrow{\hspace*{1cm}}}\ \frac{-i}{\sqrt{2}}\ket{1}\otimes\ket{+}^{\otimes n} + \ket{0}\otimes \frac{1}{\sqrt{2^{n+1}}}\paren{\ket{0}+i\ket{1}}^{\otimes n}
\end{equation*}
Next, $\mc{B}_i$ transforms the state to
\begin{equation*}
\frac{-i}{\sqrt{2}}\ket{1}\otimes\ket{+}^{\otimes n} + \ket{0}\otimes \frac{1}{\sqrt{2^{n+1}}}\paren{\ket{0}+i\ket{1}}^{\otimes n} \ \underset{\mc B_i(\pi/4)}{\xrightarrow{\hspace*{1cm}}}\ \frac{-ie^{-in\pi/4}}{\sqrt{2}}\ket{1}\otimes\ket{+}^{\otimes n} + \frac{1}{\sqrt{2}}\ket{0}\otimes\ket{0}^{\otimes n} 
\end{equation*}
and finally, the central mixing term $\mc B_0$ gives (with $\omega:= e^{-in\pi/4}$)
\begin{equation*}
\frac{-i\omega}{\sqrt{2}}\ket{1}\otimes\ket{+}^{\otimes n} + \frac{1}{\sqrt{2}}\ket{0}\otimes\ket{0}^{\otimes n} \ \underset{\mc B_0(\pi/4)}{\xrightarrow{\hspace*{1cm}}}\ \frac{-i}{2}\ket{1}\otimes\paren{\omega\ket{+}^{\otimes n} - \ket{0}^{\otimes n}} + \frac{1}{2}\ket{0}\otimes\paren{\omega\ket{+}^{\otimes n} + \ket{0}^{\otimes n}}
\end{equation*}
which is the final state $\ket{\psi_f}$. The success probability is then
\begin{equation}
  \text{Pr(success)} = |\braket{\str 0}{\psi_f}|^2 = \frac{1}{4}\left|1 -\omega\braket{0}{+}^n \right|^2 = 1/4 + O(1/2^n)
\end{equation}
which is a finite constant and may be boosted polynomially close to 1 with a logarithmic number of repetitions. 

\subsubsection{Other symmetric instances}
\label{sec:QAOAsym}

The success of QAOA1 on the two chosen instances is in part due to the fact that only the potential on the support of the initial state affects the state dynamics. This feature is absent from the other algorithms studied here. Notably, for the adiabatic algorithm on \texttt{Spike}, while it is true that the spectral gap is minimized at the same point $u^*$ as for the ramp without the spike (see \cite{Kong2017}), the size of the gap itself depends on the spike parameters, so that in particular, when the spike is sufficiently broad or tall, the gap becomes exponentially small in $n$. In stark contrast, the performance of QAOA1 is \emph{independent} of the gap parameters, since the state has vanishing support on the spike.

Now, we will use this feature to give conditions under which a symmetric cost function may be successfully minimized by QAOA1. When the cost can be decomposed into a linear ramp and a super-linear part that has small support on the initial state, one may ignore the super-linear terms and treat the problem as a linear ramp. Suppose we have a Hamming-symmetric cost function $c(\tilde w) = c_0 + c_1\tilde w + c_2\tilde w^2 + \cdots$, written as a Taylor series in $\tilde w$, the shifted Hamming weight $\tilde w = w - n/2$ (which we henceforth replace with $w$). Separate the function into a linear part and a super-linear part, $c(w) = r(w) + q(w)$, where
\begin{align}
r(w) &= c_0 + c_1 w\\
s(w) &= c_2w^2 + \cdots
\end{align}

Under Lemma~\ref{lem:support}, if it is the case that $s(w)$ overlaps weakly with the initial state (which is roughly supported on weights $n/2\pm O\paren{\sqrt{n}}$), and if the addition of $s(w)$ does not change the global minimum of $r(w)$, then such a cost function $c(w)$ can be optimized using a ``ramp protocol'' for $r(w)$, as was done for the \texttt{Spike} problem in Sec.~\ref{sec:QAOAspike} (provided the slope of the ramp $|c_1|\ge O(1/\text{poly}(n))$).

However, in this case we can do better (Theorem~\ref{thm:linearize} below): even if the global minimum of $c(w)$ does not coincide with that of $r(w)$, the ramp protocol may be suitably modified to ensure the successful minimization of $c(w)$. Suppose $\min_{w}c(w) = w^*$. For the ramp $r(w) = c_0 + c_1 w$, the first step of QAOA is evolution under $\mc C(\pi/(2|c_1|))$. For $c(w)$, we modify $\gamma$ to $\gamma^*$ (to be determined), and keep $\beta =\pi/4$ unchanged. Then, the final state may be written as
  \begin{align}
    \ket{\psi_f} &= \bigotimes_{i=1}^n\paren{\sin\paren{\gamma^*/2}\ket{0} + \cos\paren{\gamma^*/2}\ket{1}}\\
    &= \suml{w=0}{n}{\paren{\sin\paren{\gamma^*/2}}^{n-w}\paren{\cos\paren{\gamma^*/2}}^w\binom n w^{1/2}\ket{w}}
  \end{align}
Then, by inspection, $\gamma^*$ must maximize the success probability, or equivalently, the function $\paren{\sin\paren{\gamma^*/2}}^{2(n-w^*)}\paren{\cos\paren{\gamma^*/2}}^{2w^*}$. An elementary calculation yields that
\begin{equation}
  \gamma^* = \arccos\sqrt{\frac{w^*}{n}}
\end{equation}
Finally, the success probability is
\begin{equation}
  \prob \text{(success)} = \frac{(w^*)^{w^*}(n-w^*)^{w^*}}{n^n}\binom{n}{w^*} = O(1)
\end{equation}
by Stirling's approximation. So, QAOA1 with $\beta=\pi/4,\gamma=\gamma^*$ successfully optimizes the cost function $c(w)$. Finally, we note that if the minimum $w^*$ of $c$ is unknown, the above QAOA1 protocol may be carried out for all $n+1$ possible values of $w^*$ until success, which is at most a factor $O(n)$ overhead. Therefore, we have just proven the following result:
\begin{theorem}
  \label{thm:linearize}
  When $c(w) = r(w) + s(w)$ and $r$ is linear in $w$ with slope $\Omega(1/\text{poly}(n))$, and $s(w)$ satisfies the weak overlap condition 3 in Lemma~\ref{lem:support}, $c(w)$ can be successfully minimized via QAOA1 with at most a polynomial number of classical repetitions.  
\end{theorem}
There is an intuitive picture for the feature of QAOA discovered in Theorem~\ref{thm:linearize} above. As has been observed before \cite{Bringewatt2017, Brady2016}, the low energy spectrum of the mixing operator $B$ can be mapped to a suitable harmonic oscillator that treats the Hamming weight $w$ as the position variable. Under this mapping, the initial state $\ket{+}^{\otimes n}$ acts as the vacuum state wavepacket, and a linear ramp with slope $a$, $C = a\sum_w w\ket{w}\bra{w}$ is the analogous position operator. We may then qualitatively work out the action of QAOA on the initial wavepacket. The first round, evolution under $C$, displaces the vacuum to a state with finite momentum $p = a \gamma$. Then, evolution under the harmonic oscillator Hamiltonian $B$ for time $\beta=\pi/2$ rotates the coherent state so that the final state is one that is displaced in $w$. So, in a single round of QAOA, the wavepacket gains momentum and propagates to a new location in hamming weight space. (This feature has been recently noted in \cite{verdon2018}.) While the above method recovers the QAOA1 protocol qualitatively, it gets the angle $\gamma$ wrong by a factor $2/\pi$. This is due to the curvature of the phase space. In fact, the wavepacket is more accurately described by a \emph{spin-coherent} state, in which the conjugate operators are the total spin operators $S_x$ and $S_z$. It remains to be seen how this (spin-)coherent state picture may be employed to understand the behavior of QAOA on other (especially non-Hamming symmetric) instances. The simplicity of this description suggests a classical algorithm which simulates the momentum transfer and jump operations of the wavepacket via local gradient measurements of the cost function. This could give rise to a new, quantum-inspired classical search heuristic that escapes local minima more efficiently than existing classical methods.  

\noindent \textbf{Acknowledgments:} We thank P.~S.~Krishnaprasad for helpful discussions. This work was supported in part by the U.S. Department of Energy, Office of Science, Office of Advanced Scientiﬁc Computing Research, Quantum Algorithms Teams program. A.~B.\ acknowledges support from the QuICS Lanczos Graduate Fellowship. 
%\section{Conclusion}
%\label{sec:conclusion}

\begin{appendices}
\section{The control framework}
\label{sec:control}

Given a dynamical equation depending on additional parameters (which we call the \emph{controls}), what properties does a control protocol which optimizes a given cost function satisfy? The relevance of this question extends across many fields where optimal control (with respect to a cost function) is desired. In fact, it has been observed \cite{mcclean2016, Yang2017} that the optimal control problem also applies to heuristic optimization algorithms, where the controlled dynamics are described precisely by Schr\"{o}dinger evolution under the annealing Hamiltonian, and the cost function is given by the energy of the final state. 

Consider a first-order differential equation describing the dynamics of an $n$-dimensional real vector $x\in \mbb{R}^n$, and controlled by $m$ control parameters which we denote by the vector $u\in\mbb{R}^m$:
\begin{equation}
  \dot{x}(t) = f(x(t),u(t))
\end{equation}

The functional form $f$ may be very general; we only assume that $f$ is ``Markovian'' (i.e., depends only on the current state $(x(t),u(t))$), and that there is no explicit time-dependence. Typically, it is further assumed that the control $u$ inhabits a fixed, \emph{compact} subset, $u\in \mc{U}\subset\mbb{R}^m$. The domain $\mc{U}$ represents a feasible set of controls.  

In order to talk about optimal control, we must first specify a notion of cost. In a real problem such as optimizing the trajectory of a spacecraft, the cost might be expressed in terms of time, amount of fuel used (i.e. a trajectory-dependent cost), and the distance of the final position from the target location (i.e., a final state cost). Thus, the cost function may generally be expressed as a (weighted) sum of three costs:
\begin{enumerate}
\item the total time for the process, $T = \intl{0}{T}{1\text{d}t}$
\item the \emph{running} cost, which is given as an integral over the running time, $\intl{0}{T}{L\paren{x(t),u(t),t})\text{d}t}$
\item the \emph{terminal} cost, which is a final state-dependent function $K(x(T))$. 
\end{enumerate}

The full cost function may be expressed in the general form
\begin{equation}
\label{eq:inftime}
  J = K(x_{final}) + \intl{0}{\infty}{L\paren{x(t),u(t),t}\text{d}t}
\end{equation}
where $J$ is a functional of the control schedule $u(t)$ and the dynamical path $x(t)$.
The objective is to find the control function $u(t)$, over all piecewise continuous functions $u:\mathbb R_{\ge 0}\rightarrow \mc{U}$, that minimize the overall cost, i.e. $\arg\min_{u(t)}J(u)$. This is the so-called infinite time horizon formulation of the problem. Alternatively, one can fix the total time for the protocol $T$ to be finite. Then, we are asked to minimize over all piecewise continuous functions $u:[0,T]\rightarrow \mc{U}$ the cost
\begin{equation}
\label{eq:inftime}
J = K(x(T)) + \intl{0}{T}{L\paren{x(t),u(t),t}\text{d}t}
\end{equation}

A wealth of literature in classical control theory discusses the question of optimal control, and we emphasize its potential applicability in the setting of designing efficient heuristic optimizers, both classical and quantum. Here, we will focus on one result, the \emph{Pontryagin Minimum Principle} (PMP), which imposes necessary conditions for a control protocol to be optimal using the so-called control Hamiltonian description.  

The control Hamiltonian $\mc H$ is a classical functional describing auxiliary Hamiltonian dynamics on a set of variables given by $x$ and corresponding \emph{co-state} (or conjugate momentum) variables $p$. The conjugate momenta depend on the cost function $J$ in Eq.~\ref{eq:inftime}, and are introduced as Lagrange multipliers that impose the equations of motion for each coordinate of $x$. The full cost function (at time $t$), which includes the cost terms in $J$ and the constraints, is given by the control Hamiltonian $\mc{H}$.
\begin{equation}
  \mc{H} := L(x,u) - p\cdot f(x,u)
\end{equation}
Then, PMP states that the optimal control is one which minimizes the control Hamiltonian at all times. That is,
\begin{equation}
\label{eq:pmp}
\mc{H}\paren{x(t),p(t),u^{*}} \le \min_{u\in \mc U}\mc{H}\paren{x(t),p(t),u}
\end{equation}

In the special case when $\mc H$ is linear in the control $u$, the above minimality condition is satisfied only if the control lies on the boundary of the feasible set $\mc U$. This implies that optimal trajectories are bang-bang, i.e., the controls only take their extremal values. The optimal point(s) on the boundary are determined by the intersection of the constant-$\mc H$ hyperplanes in control space with the set boundary. However, an important exception arises when the derivative of $\mc H$ with respect to $u$ vanishes over a finite interval. In this case, the control becomes singular, i.e., its optimal value no long lies solely on the boundary of $\mc U$. 

The control framework described here covers many heuristic optimization algorithms, and we will fix some notation to suit this setting. The dynamical vector of interest will a state $\ket{\psi}$ (quantum) or $\cket{\psi}$ (classical), and the generator of dynamics will be a \emph{controlled} linear operator  
\begin{equation}
  H(\mb{u}) = \suml{i=0}{m}{u_i H_i} \equiv \mb{u}\cdot\mb{H}
\label{eq:ham}
\end{equation}

where $\mb{u}$ and $\mb{H}$ are vectors with components $u_i$ and $H_i$ respectively. We assume that individual Hamiltonians $H_i$ are time-independent, and only their overall strength, controlled by the coefficient $u_i(t)$, is time-dependent. We will fix the range of all $u_i$ to $[0,1]$.
\section{Bang-bang simulated annealing on the \texttt{Spike}}
\label{sec:BBSAspikeApp}
For \texttt{Spike}, the strategy used for \texttt{Bush}, namely, run randomized gradient descent (zero-temperature SA) from start to finish, fails due to the presence of a barrier. So, if we run gradient descent for time $O(n)$ per walker, then we are left with a distribution sharply peaked at the false minimum. We may now attempt to diffuse across the barrier. For a sufficiently wide barrier, this strategy will again fail, since the diffusion rate across the spike is exponentially small in $n^{a}$. However, we instead turn on diffusion for a short time, so that a constant fraction of the walkers ``hop on'' the barrier, while the rest diffuse away from the barrier. Then, we turn on randomized gradient descent again until the finish. The fraction of walkers on the barrier are now guaranteed to walk to the global minimum in time $O(n)$, as the slope is positive.

So, it can be seen that for the spike problem, an algorithm with the same structure as SA but a schedule that is designed without the adiabaticity constraint, successfully finds the global minimum, and thus exponentially outperforms SA (and QAO for certain parameter regimes, see Table~\ref{tab:results}) on the same instance. It should be noted that the success of BBSA depends sensitively on the shape of the spike. In particular, we expect success (i.e. at least $1/\text{poly}(n)$ walkers reach the global minimum) when the part of the spike with positive slope (i.e. the ``uphill'' portion) has width $O(\log n)$.

\section{Proof of Lemma~\ref{lem:support}}
\label{sec:appC}

Let $\mc C = e^{-i\gamma \sum_w{c(w)\ket{w}\bra{w}}}$ and let $\mc R, \mc S$ be defined analogously with the cost terms $r(w)$ and $s(w)$, where $c(w) = r(w) + s(w)$. $\mc R$ and $\mc S$ are mutually commuting, so $\mc C = \mc R\mc S$, and the first step of the QAOA1 protocol may be written as
  \begin{equation}
    \mc C\ket{\psi_0} = \mc R\mc S\ket{\psi_0} = \mc R\suml{w=0}{n}{e^{-i\gamma s(w)}A_w\ket{w}} = \mc R\ket{\psi_0} + \mc R\suml{w=0}{n}{\paren{e^{-i\gamma s(w)}-1}A_w\ket{w}}
  \end{equation}
  After the mixing operator $\mc B = e^{-i\beta B}$ is applied, the final state is
  \begin{equation}
    \ket{\psi_f} = \mc B\mc R\ket{\psi_0} + \mc B\mc R\suml{w=0}{n}{\paren{e^{-i\gamma s(w)}-1}A_w\ket{w}}
  \end{equation}
  The overlap with the global minimum $\ket{\psi^*}$ is
  \begin{align}
    \braket{\psi^*}{\psi_f} &= \bra{\psi^*}\mc B\mc R\ket{\psi_0} + \bra{\psi^*}\mc B\mc R\suml{w=0}{n}{\paren{e^{-i\gamma s(w)}-1}A_w\ket{w}}\\\label{eq:lem}
    \implies |\braket{\psi^*}{\psi_f} - \bra{\psi^*}\mc B\mc R\ket{\psi_0}| &=  |\bra{\psi^*}\mc B\mc R\suml{w=0}{n}{2e^{i\gamma s(w)/2 - i\pi/2}\sin\paren{\frac{\gamma s(w)}{2}}A_w\ket{w}}
  \end{align}
Now, $p = |\bra{\psi^*}\mc B\mc R\ket{\psi_0}|^2$, and let $p^* = |\braket{\psi^*}{\psi_f}|^2$, the success probabilities of QAOA1 on $r(w)$ and the full cost function $c(w)$, respectively. We wish to show that $p^*$ is at least $p - o(p)$. Using the triangle inequality $|x| - |y| \le |x-y|$ on the left side of Eq.~\ref{eq:lem}, and Cauchy-Schwarz inequality $|\braket u v | \le |\braket u u|^{1/2}|\braket v v|^{1/2}$ on the right side, we get the following:
  \begin{align}
    \sqrt p  - \sqrt{p^*} &\le \suml{w=1}{n}{4|A_w|^2\sin^2\paren{\frac{\gamma s(w)}{2}}}^{1/2} = \sqrt q \\
    \implies p^* &\ge p\paren{1- \sqrt{q/p}}^2 = p\paren{1- o(1)}^2 = p - o(p)
  \end{align}
which proves the lemma.
\end{appendices}

\bibliographystyle{unsrt}
\bibliography{library}

\begin{thebibliography}{10}

\bibitem{mcclean2016}
J.~R. McClean, J.~Romero, R.~Babbush, and A.~Aspuru-Guzik.
\newblock The theory of variational hybrid quantum-classical algorithms.
\newblock {\em New Journal of Physics}, 18(2):023023, 2016.

\bibitem{Yang2017}
Z.~C. Yang, A.~Rahmani, A.~Shabani, H.~Neven, and C.~Chamon.
\newblock {Optimizing variational quantum algorithms using pontryagin's minimum
  principle}.
\newblock {\em Physical Review X}, 7(2):1--8, 2017.

\bibitem{nisq}
J.~Preskill.
\newblock Quantum computing in the {NISQ} era and beyond.
\newblock {\em arXiv preprint arXiv:1801.00862}, 2018.

\bibitem{Farhi2000}
E.~Farhi, J.~Goldstone, S.~Gutmann, and M.~Sipser.
\newblock Quantum computation by adiabatic evolution.
\newblock {\em arXiv preprint quant-ph/0001106}, 2000.

\bibitem{Farhi2014}
E.~Farhi, J.~Goldstone, and S.~Gutmann.
\newblock A quantum approximate optimization algorithm.
\newblock {\em arXiv preprint arXiv:1411.4028}, 2014.

\bibitem{hogg2000}
T.~Hogg and D.~Portnov.
\newblock Quantum optimization.
\newblock {\em Information Sciences}, 128(3-4):181--197, 2000.

\bibitem{hadfield2017}
S.~Hadfield, Z.~Wang, B.~O'Gorman, E.~G. Rieffel, D.~Venturelli, and R.~Biswas.
\newblock From the quantum approximate optimization algorithm to a quantum
  alternating operator ansatz.
\newblock {\em arXiv preprint arXiv:1709.03489}, 2017.

\bibitem{Farhi2018}
E.~Farhi and H.~Neven.
\newblock Classification with quantum neural networks on near term processors.
\newblock {\em arXiv preprint arXiv:1802.06002}, 2018.

\bibitem{Kim2017}
I.~H. Kim and B.~Swingle.
\newblock Robust entanglement renormalization on a noisy quantum computer.
\newblock {\em arXiv preprint arXiv:1711.07500}, 2017.

\bibitem{wecker2016}
D.~Wecker, M.~B. Hastings, and M.~Troyer.
\newblock Training a quantum optimizer.
\newblock {\em Physical Review A}, 94(2):022309, 2016.

\bibitem{verdon2018}
G.~Verdon, J.~Pye, and M.~Broughton.
\newblock A universal training algorithm for quantum deep learning.
\newblock {\em arXiv preprint arXiv:1806.09729}, 2018.

\bibitem{pontryagin}
L.~S. Pontryagin.
\newblock {\em Mathematical theory of optimal processes}.
\newblock Routledge, 2018.

\bibitem{Farhi2002}
E.~Farhi, J.~Goldstone, and S.~Gutmann.
\newblock Quantum adiabatic evolution algorithms versus simulated annealing.
\newblock {\em arXiv preprint quant-ph/0201031}, 2002.

\bibitem{Brady2016}
L.~T. Brady and W.~van Dam.
\newblock Spectral-gap analysis for efficient tunneling in quantum adiabatic
  optimization.
\newblock {\em Physical Review A}, 94(3):032309, 2016.

\bibitem{mitra1986}
D.~Mitra, F.~Romeo, and A.~Sangiovanni-Vincentelli.
\newblock Convergence and finite-time behavior of simulated annealing.
\newblock {\em Advances in applied probability}, 18(3):747--771, 1986.

\bibitem{geman1984}
S.~Geman and D.~Geman.
\newblock Stochastic relaxation, gibbs distributions, and the bayesian
  restoration of images.
\newblock {\em IEEE Transactions on pattern analysis and machine intelligence},
  (6):721--741, 1984.

\bibitem{hajek1988}
B.~Hajek.
\newblock Cooling schedules for optimal annealing.
\newblock {\em Mathematics of operations research}, 13(2):311--329, 1988.

\bibitem{gidas1985}
B.~Gidas.
\newblock Nonstationary markov chains and convergence of the annealing
  algorithm.
\newblock {\em Journal of Statistical Physics}, 39(1-2):73--131, 1985.

\bibitem{jansen2007}
S.~Jansen, M.-B. Ruskai, and R.~Seiler.
\newblock Bounds for the adiabatic approximation with applications to quantum
  computation.
\newblock {\em Journal of Mathematical Physics}, 48(10):102111, 2007.

\bibitem{elgart2012}
A.~Elgart and G.~A. Hagedorn.
\newblock A note on the switching adiabatic theorem.
\newblock {\em Journal of Mathematical Physics}, 53(10):102202, 2012.

\bibitem{roland2002}
J.~Roland and N.~J. Cerf.
\newblock Quantum search by local adiabatic evolution.
\newblock {\em Physical Review A}, 65(4):042308, 2002.

\bibitem{somma2012}
R.~D. Somma, D.~Nagaj, and M.~Kieferov{\'a}.
\newblock Quantum speedup by quantum annealing.
\newblock {\em Physical review letters}, 109(5):050501, 2012.

\bibitem{chakrabarti2007}
R.~Chakrabarti and H.~Rabitz.
\newblock Quantum control landscapes.
\newblock {\em International Reviews in Physical Chemistry}, 26(4):671--735,
  2007.

\bibitem{rahmani2011}
Armin Rahmani and Claudio Chamon.
\newblock Optimal control for unitary preparation of many-body states:
  Application to luttinger liquids.
\newblock {\em Physical review letters}, 107(1):016402, 2011.

\bibitem{peruzzo2014}
A.~Peruzzo, J.~McClean, P.~Shadbolt, M.-H. Yung, X.-Q. Zhou, P.~J. Love,
  A.~Aspuru-Guzik, and J.~L. O’brien.
\newblock A variational eigenvalue solver on a photonic quantum processor.
\newblock {\em Nature communications}, 5:4213, 2014.

\bibitem{lapert2011}
M.~Lapert, Y.~Zhang, S.~J. Glaser, and D.~Sugny.
\newblock Towards the time-optimal control of dissipative spin-1/2 particles in
  nuclear magnetic resonance.
\newblock {\em Journal of Physics B: Atomic, Molecular and Optical Physics},
  44(15):154014, 2011.

\bibitem{bonnard2012}
B.~Bonnard, S.~J. Glaser, and D.~Sugny.
\newblock A review of geometric optimal control for quantum systems in nuclear
  magnetic resonance.
\newblock {\em Advances in Mathematical Physics}, 2012, 2012.

\bibitem{Kong2017}
L.~Kong and E.~Crosson.
\newblock The performance of the quantum adiabatic algorithm on spike
  hamiltonians.
\newblock {\em International Journal of Quantum Information}, 15(02):1750011,
  2017.

\bibitem{Reichardt2004}
B.~W. Reichardt.
\newblock The quantum adiabatic optimization algorithm and local minima.
\newblock In {\em Proceedings of the Thirty-sixth Annual ACM Symposium on
  Theory of Computing}, STOC '04, pages 502--510, New York, NY, USA, 2004. ACM.

\bibitem{Crosson2016}
E.~Crosson and A.~W. Harrow.
\newblock Simulated quantum annealing can be exponentially faster than
  classical simulated annealing.
\newblock In {\em Foundations of Computer Science (FOCS), 2016 IEEE 57th Annual
  Symposium on}, pages 714--723. IEEE, 2016.

\bibitem{Bringewatt2017}
J.~Bringewatt, W.~Dorland, S.~P. Jordan, and A.~Mink.
\newblock Diffusion monte carlo versus adiabatic computation for local
  hamiltonians.
\newblock {\em arXiv preprint arXiv:1709.03971}, 2017.

\end{thebibliography}

\end{document}